\documentclass[smallcondensed]{svjour3}
\usepackage{epsfig}
\usepackage{graphicx}

\usepackage{amsmath}
\usepackage{amsfonts}
\usepackage{natbib}

\newcommand{\rr}{\mathbf{R}}

\newcommand{\zz}{\mathbf{Z}}

\def\figfig#1#2#3{\centerline{\epsfig{figure=#2.eps,height=#1}}%
\caption{\small #3}\label{fig:#2}}

\journalname{CMDA}

\begin{document}
\title{Shadowing Lemma and Chaotic Orbit Determination}
\author{Federica Spoto         \and
        Andrea Milani
}
\institute{F. Spoto \and A. Milani \at
              Dipartimento di Matematica, Universit\`a di Pisa,
              Largo Pontecorvo 5,
              56127 Pisa, Italy
              \email{spoto@spacedys.com}
} 
\date{5 June 2015}
\maketitle

\begin{abstract}
Orbit determination is possible for a chaotic orbit of a dynamical
system, given a finite set of observations, provided the initial
conditions are at the central time. In a simple discrete model, the
standard map, we tackle the problem of chaotic orbit determination
when observations extend beyond the predictability horizon. If the
orbit is hyperbolic, a shadowing orbit is computed by the least
squares orbit determination. We test both the convergence of the orbit
determination iterative procedure and the behaviour of the
uncertainties as a function of the maximum number $n$ of map
iterations observed.  When the initial conditions belong to a chaotic
orbit, the orbit determination is made impossible by numerical
instability beyond a computability horizon, which can be approximately
predicted by a simple formula. Moreover, the uncertainty of the
results is sharply increased if a dynamical parameter is added to the
initial conditions as parameter to be estimated. The uncertainty of
the dynamical parameter decreases like $n^a$ with $a<0$ but not large
(of the order of unity). If only the initial conditions are estimated,
their uncertainty decreases exponentially with $n$.  If they belong to
a non-chaotic orbit the computational horizon is much larger, if it
exists at all, and the decrease of the uncertainty is polynomial in
all parameters, like $n^a$ with $a\simeq 1/2$. The Shadowing Lemma
does not dictate what the asymptotic behaviour of the uncertainties
should be. These phenomena have significant implications, which remain
to be studied, in practical problems of orbit determination involving
chaos, such as the chaotic rotation state of a celestial body and a
chaotic orbit of a planet-crossing asteroid undergoing many close
approaches.

\end{abstract}

\section{Introduction}
\label{sec:intro}

Chaotic dynamical systems are characterized by the existence of a
\textit{predictability horizon} in time, beyond which the information
on the state available from the initial conditions is not enough for
meaningful predictions.  Thus it appears a difficult task to perform
an orbit determination for a chaotic dynamical system, at least when
the observations are spread over a time-span longer than the
predictability horizon.

Nevertheless there are practical problems of orbit determination in
which the system is chaotic and the time-span of the
  observations is very long. It is important to understand the
behaviour of the solutions, with their estimated uncertainties, in
particular when the variables to be solved for include not
just the initial conditions but also some dynamical parameters. If the
number of available observations grows, but simultaneously
the time interval over which they are spread grows up to
values comparable to the predictability horizon, does the solution
become more accurate, and is the iterative procedure of
  differential corrections \citep[Chap. 5]{orbit_det} to find the least
  squares solution still possible?

In this paper we use a model problem, namely the discrete dynamical
system defined by the standard map of the pendulum, with just one
dynamical parameter, the $\mu$ coefficient appearing in
equation~(\ref{stmap}). We also set up an observation process
  in which both coordinates of the standard map are observed after
  each iteration. In the observations we include a simulated random
noise with a normal distribution. Then, each experiment of
orbit determination is also a concrete computation of a segment
limited to $n$ iterations (of the map and of its inverse) of a
\textit{$\varepsilon$ shadowing orbit} for the \textit{$\delta$ pseudo
  trajectory} defined by the observation process. The
\textit{Shadowing Lemma} (see Section~\ref{sec:SL}) provides a
mathematically rigorous result on the availability of shadowing
orbits, but thanks to the orbit determination process we make explicit
the relationship between $\varepsilon$ and $\delta$ (see
Section~\ref{sec:SL_OD}), and we explicitly compute the
$\varepsilon$-shadowing orbit.

At the same time, each experiment provides an estimate of the standard
deviation of each of the variables, including initial
conditions and the parameter. These estimates can be plotted as a
function of $n$, thus showing the relationship between accuracy,
number of observations and time interval, measured in Lyapounov times
(see Section~\ref{sec:numerical}).

Of course the numerical experiments are limited to a finite number of
iterations, while the Shadowing Lemma refers to an infinite
orbit. However, the maximum number of iterations is
controlled by another time limit, the \textit{computability horizon}
due to round off error. This limit can be estimated approximately by a
simple formula, and it is found in numerical experiments as a function
of both the initial conditions and the numeric precision used in the
computations.

\subsection{Wisdom hypothesis}
\label{sec:wisdom}

In 1987 J. Wisdom was discussing the chaotic rotation state of
Hyperion, when he claimed that numerical experiments indicated that
\textit{the knowledge gained from measurements on a chaotic dynamical
  system grows exponentially with the timespan covered by the
  observations} \citep{wisdom87}. This pertained in particular
to the information on dynamical parameters like the moments of inertia
ratios for Hyperion, as well as the rotation state at the midpoint of
the time interval covered by the observations, which he proposed would
be determined with exponentially decreasing uncertainty.

Therefore Wisdom suggests that the orbit determination for a
chaotic system might be in fact more effective than for a non-chaotic
one.  It is clear from the context that he was referring to
numerical results, thus his statement can only be verified with finite
computations as close as possible to a realistic data processing of
observations of a chaotic system with dynamical parameters to be
determined.

We have set as a goal in this paper to test the behavior of
the uncertainty in the dynamical parameter of our model
  problem. We shall discuss the implications for Wisdom's claim in
Section~\ref{sec:discwisd}.

\subsection{Application to planet-crossing asteroids}

In our solar system there are \textit{planet-crossing minor bodies},
including asteroids and comets, by definition such that their
  orbits can, at some times, intersect the orbit of the major
planets. In particular many of the \textit{Near Earth Asteroids
  (NEAs)} can intersect the orbit of the Earth. These orbits are
necessarily chaotic, at least over the timespan accessible to accurate
numerical computations.

Unfortunately, these orbits are especially important and necessary to
be studied because of the very reason of chaos, namely close
approaches to the major planets including the Earth: these approaches
may, in some cases, be actual impacts on a finite size planet. 

The attempt to predict possibility of impacts by NEAs, in particular
on our planet, is called \textit{Impact Monitoring}, and it is indeed
a form of orbit determination for chaotic orbits. There is a subset of
cases of NEAs for which non-gravitational perturbations, such as the
ones resulting from the Yarkovsky effect, are not negligible
in terms of Impact Monitoring because of the exponential
divergence of nearby orbits which amplifies initially very small
perturbations\\ \citep{apophis,bennu,1950DA,2009FD}.

Thus the Impact Monitoring for these especially difficult cases is an
instance of orbit determination of a chaotic system, with as
  parameters the 6 initial conditions and at least one dynamical
  parameter, such as a Yarkovsky effect coefficient to be solved for.
We shall show in Section~\ref{sec:impacts} that the weak determination
of the dynamical parameter is a key feature of these cases.

\section{Orbit determination for the standard map}
\label{sec:OD_stmap}
The simplest example of a conservative dynamical system which has both
chaotic and ordered orbits can be built by means of an area preserving
map of a two dimensional manifold:
\begin{equation}\label{stmap}
  S_{\mu}(x_k,y_k)=\left\{
    \begin{array}{ccl}
      x_{k+1} & = & x_k+y_{k+1}\\
      y_{k+1} & = & y_k-\mu \sin{x_k}
    \end{array}
  \right.
\end{equation}
where $\mu$ is the perturbation parameter, and $S$ is the standard
map. The system has more regular orbits for small $\mu$, and
more chaotic orbits for large $\mu$. We choose an
intermediate value of $\mu$, e.g. $\mu=0.5$, in such a way that
ordered and chaotic orbits are both
present. Figure~\ref{fig:smap_05_zoom} shows the strongly chaotic
region around the hyperbolic fixed point, and a few regular orbits on
both sides.
\begin{figure}[h!]
  \figfig{8cm}{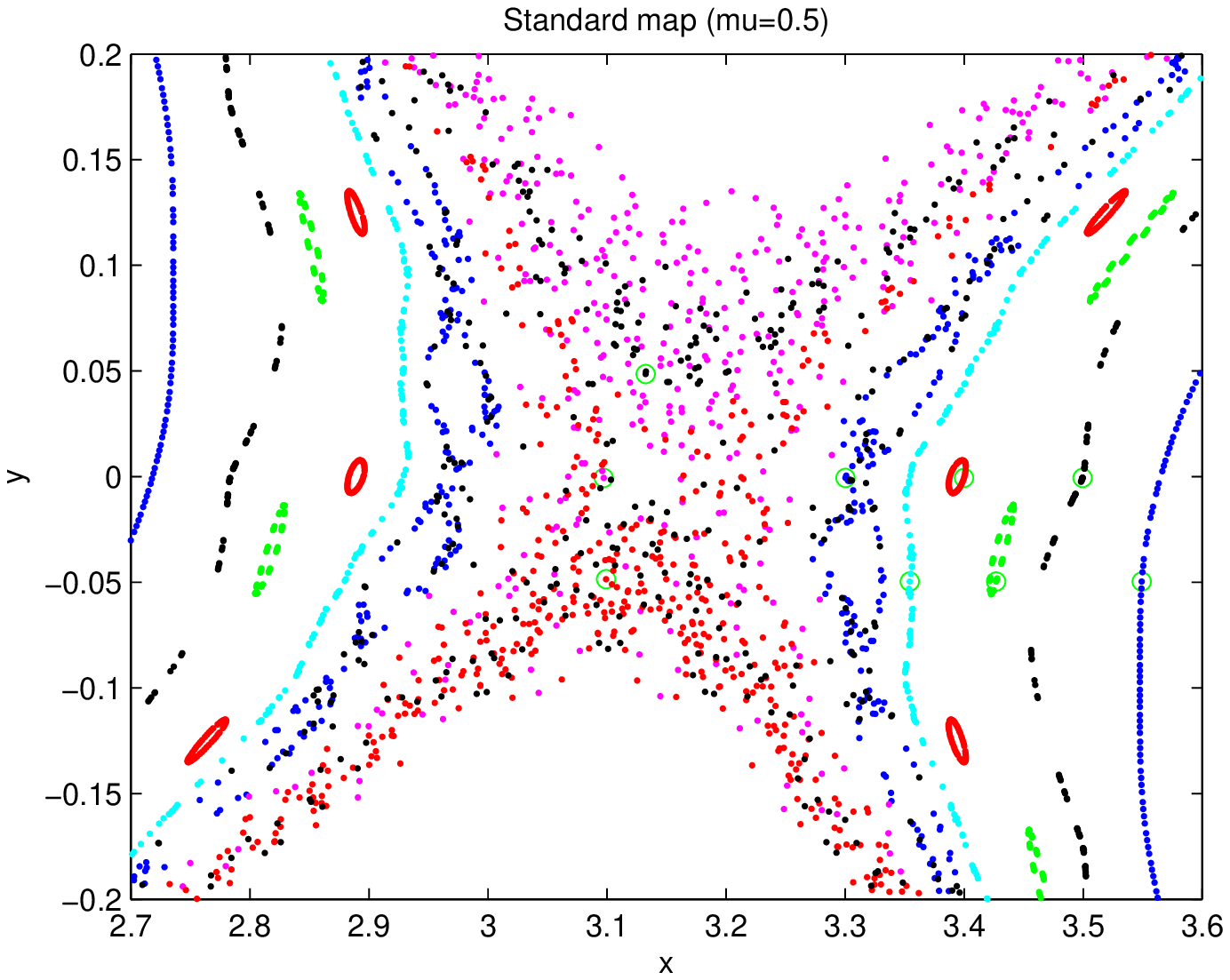}{Orbits of the standard map for the
    perturbation parameter $\mu=0.5$. Plotted is a blow up of
    the central region around the hyperbolic fixed point, showing the
    strongly chaotic region and a few regular orbits on both sides.}
  \label{smap_zoom}
\end{figure}

The advantage of such example is that the least square parameter
estimation process can be performed by means of an explicit formula.

First we compute the linearized map
\begin{displaymath}
  DS = \left(
  \begin{array}{cc}
    \frac{\partial x_{k+1}}{x_{k}} & \frac{\partial x_{k+1}}{y_{k}}  \\
    \frac{\partial y_{k+1}}{x_{k}} & \frac{\partial y_{k+1}}{y_{k}}
  \end{array}
  \right)
  = \left(
  \begin{array}{cc}
    1-\mu \cos(x_{k}) & 1  \\
    -\mu \cos(x_{k}) & 1
  \end{array}
  \right)
\end{displaymath}
and from this the linearized state transition matrix
\begin{displaymath}\label{transition_matrix}
  A_{k}=\frac{\partial(x_{k}, \ y_{k})}{\partial (x_{0}, \ y_{0})}
\end{displaymath}
which is the solution of the variational equation (for infinitesimal
displacement in the initial conditions), and given
by the recursion:
\begin{displaymath}
  A_{k+1}=DS \ A_{k}\ \ ;\ \ A_0=I\ .
\end{displaymath}
The variational equation for the derivatives with respect to the model
parameter $\mu$ is:
\begin{eqnarray*}
  \frac{\partial (x_{k+1}, \ y_{k+1})}{\partial \mu} & = & DS \ \frac{\partial (x_k, \ y_k)}{\partial         										      \mu}+\frac{\partial S_\mu}{\partial \mu}\\
  & = & DS \ \frac{\partial (x_k, \ y_k)}{\partial \mu} + 
  \left(
  \begin{array}{c}
    -\sin(x_{k}) \\
    -\sin(x_{k}) 
  \end{array}
  \right)
\end{eqnarray*}
Then we set up an observation process, in which both coordinates $x$
and $y$ are observed at each iteration, and the observations are
Gaussian random variables with mean $x_k$ ($y_k$ respectively) and
standard deviation $\sigma$: we use the notation $x_k(\mu_0,\sigma)$
to indicate that the probability density function of the observation
$x_k$ is the normal $\mathcal{N}(x_k(\mu_0), \sigma^2)$ one,
and similarly for $y_k$. The residuals are:
\begin{eqnarray}\label{residuals}
  \Bigg\{
  \begin{array}{ccl}
    \xi_k & = & x_k(\mu_0,\sigma)-x_k(\mu_1)\\
    \bar{\xi}_{k} & = & y_k(\mu_{0},\sigma)-y_{k}(\mu_1). 
  \end{array}
\end{eqnarray}
for $k=-n,\dots,n$. In (\ref{residuals}) $x_k(\mu_0,\sigma)$ and
$y_k(\mu_0,\sigma)$ are the observations at each iteration, $\mu_0$ is
the ``true'' value and $\mu_1=\mu_0+d\mu$ is the current guess.

Then the least squares fit is obtained from the normal equations
\citep{orbit_det}:
\begin{equation}\label{eq:normaleq}
  C =  \sum_{k=-n}^{n}B_{k}^{T}B_{k} \ \ ; \ \ 
  D = -\sum_{k=-n}^{n}B_{k}^{T} \left(
  \begin{array}{c}
    \xi_{k} \\
    \bar{\xi_{k}} 
  \end{array}
  \right)
\end{equation}
\[
  B_{k}=\frac{\partial({\xi_{k},\bar{\xi}_{k}})}{\partial(x_{0},y_{0},\mu)}=-\left(A_{k}|\frac{\partial(x_{k},y_{k})}{\partial
    \mu}\right).
\]
The least squares solution for both, the parameter $\mu$ and the initial conditions, is:
\[
  \left(
  \begin{array}{c}
    \delta x \\
    \delta y \\
    \delta \mu 
  \end{array}
  \right)=\Gamma D, \quad \Gamma=C^{-1}
\]
with $\Gamma$ the covariance matrix of the result. This is enough to
find the least squares solution by iteration of the
above \textit{differential correction}. However, to assess
the uncertainty of the result, weights should be assigned to
the residuals consistently with the probabilistic model, in this case
each residual needs to be divided by its standard deviation $\sigma$;
then the distribution of the result $(x,y,\mu)$ is a normal
distribution with covariance matrix $\sigma^2\,\Gamma$.

\section{Shadowing Lemma}
\label{sec:SL}

The shadowing problem is that of finding a deterministic orbit as
close as possible to a given noisy orbit. The so-called Shadowing
Lemma is the main result about shadowing near a hyperbolic set of a
diffeomorphism. Anosov~\citep{anosov} and Bowen ~\citep{bowen}
proved the existence of arbitrarily long shadowing solutions
for invertible hyperbolic maps.  Here we give an overview of
these classical results, as in \citep{pilyugin}.

Let $(X,d)$ be a metric space and let $\Phi$ be a homeomorphism
mapping $X$ onto itself.  A $\delta$-pseudotrajectory of the dynamical
system $\Phi$ is a sequence of points $\zeta=\{x_k \in X: \ k \in
\zz\}$ such that the following inequalities
\begin{eqnarray}\label{pseudotrajectory}
  d(\Phi(x_k),x_{k+1})<\delta.
\end{eqnarray}
hold. For a graphical description of a $\delta$-pseudotrajectory, see
Fig.~\ref{fig:delta_pseudo}.
\begin{figure}[h!]
  \figfig{2cm}{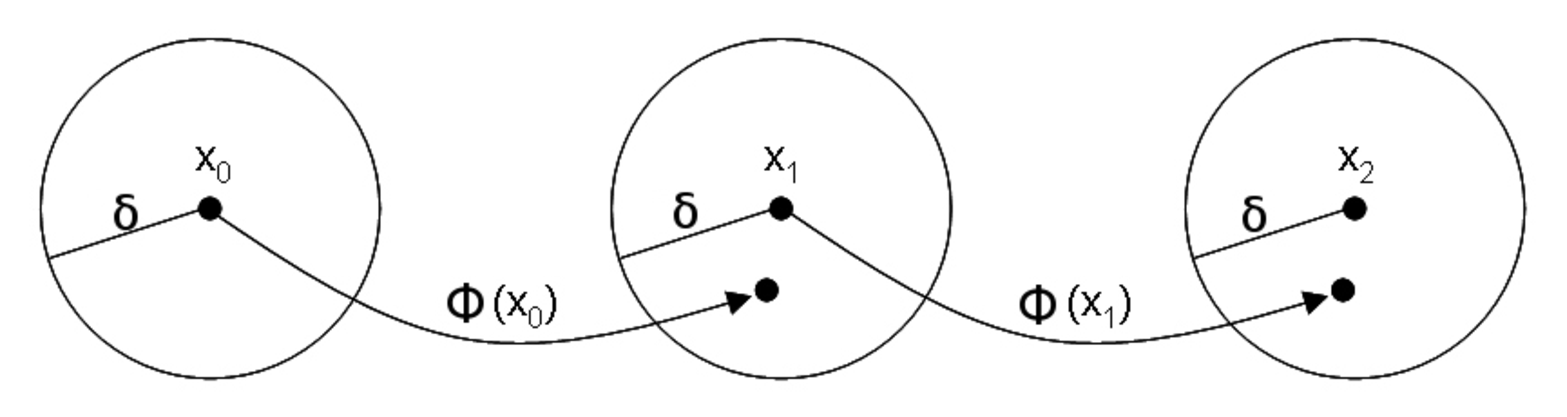}{A $\delta$-pseudotrajectory.}
  \label{delta_pseudo}
\end{figure}
Usually, a $\delta$-pseudotrajectory is considered as the result of
using a numerical method to compute orbits of the dynamical system
$\Phi$, e.g., because of round off error.  We say that a point $x \in
X$ \emph{$(\varepsilon, \Phi)$-shadows} a pseudotrajectory
$\zeta=\{x_k\}$ if the inequalities
\begin{eqnarray}\label{shadows}
  d(\Phi^k(x),x_k)<\varepsilon
\end{eqnarray}
hold (see Figure~\ref{fig:epsi_shadowing}).
\begin{figure}[h!]
  \figfig{2cm}{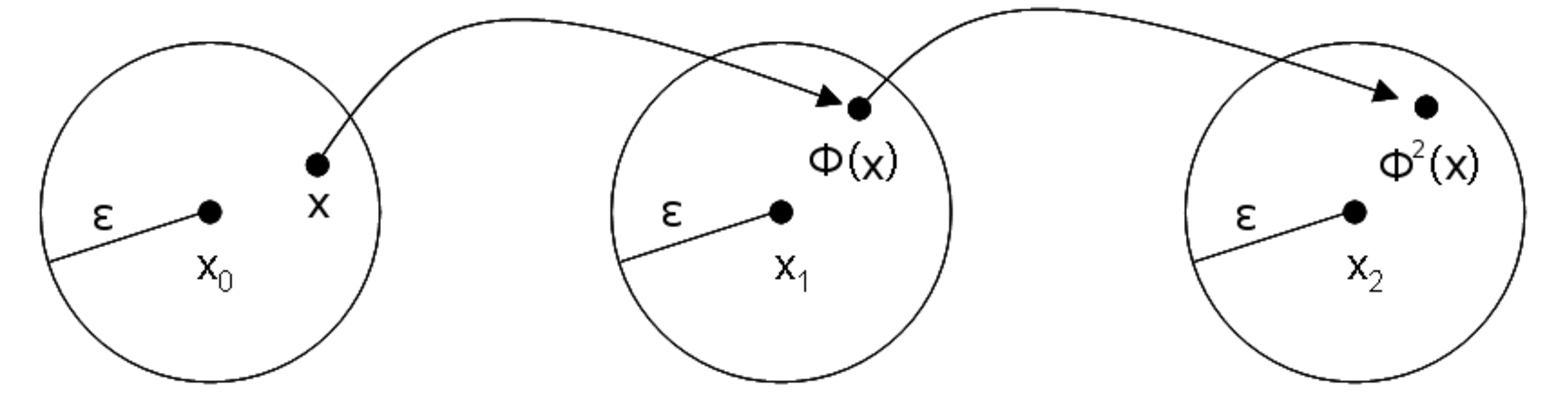}{An $\varepsilon$-shadowing.}
  \label{epsi_shadowing}
\end{figure}
If only one dynamical system $\Phi$ is considered, we will
usually write $\varepsilon$-shadows $\zeta$. The existence of a
shadowing point for a pseudotrajectory $\zeta$ means that $\zeta$ is close
to a real trajectory of $\Phi$.

The following statement is usually called the Shadowing Lemma.

\begin{theorem}
  If $\Lambda$ is a hyperbolic set for a diffeomorphism $\Phi$, then
  there exists a neighborhood $W$ of $\Lambda$ such that for all
  $\varepsilon>0$ there exists $\delta >0$ such that for any
  $\delta$-pseudotrajectory with initial conditions $\zeta\in W$ there
  is a point $x$ that $\varepsilon$-shadows $\zeta$.
\end{theorem}

The Anosov shadowing requires the existence of a hyperbolic set. It
means that at each point there are two directions where the motion is
either exponentially expanding (stable manifold) or exponentially
contracting (unstable manifold). 

\begin{definition}
We say that a set $\Lambda$ is hyperbolic for a diffeomorphism $\Phi
\in C^1(\rr^n)$ if:
\begin{itemize}
\item[(a)] $\Lambda$ is compact and $\Phi$-invariant;
\item[(b)] there exist constants $C>0$, $\lambda_0 \in (0,1)$, and families
  of linear subspaces $S(p)$, $U(p)$ of $\rr^n$, $p \in \Lambda$, such
  that
  \begin{itemize}
  \item[(b.1)] $S(p) \oplus U(p)=\rr^n$;
  \item[(b.2)] $D\Phi(p)T(p)=T(\Phi(p))$, $p \in \Lambda$, $T=S,U$;
  \item[(b.3)] 
    \begin{eqnarray*}
      |D\Phi^m(p)v| \leq C \lambda_0^m|v| \ for \ v \in S(p), \ m \geq 0 ;\\
      |D\Phi^{-m}(p)v| \leq C \lambda_0^m|v| \ for \ v \in U(p), \ m \geq 0 ;
    \end{eqnarray*}
  \end{itemize}
\end{itemize}

The definition of a hyperbolic set is strictly related to the one of
Lyapounov exponent: for each orbit inside a hyperbolic set, the
Lyapounov exponents must be either $>\log(\lambda_0)$ or
$<-\log(\lambda_0)$.
\end{definition}

\subsection{Shadowing Lemma and orbit determination}
\label{sec:SL_OD}

Our goal is to connect the Shadowing Lemma with the chaotic orbit
determination, involving the least squares fit and the differential
corrections.

First of all we build a $\delta$-pseudotrajectory by using a simulated
observations process. In Section~\ref{sec:OD_stmap} we have
set up such an observations process, with observations
$(x_k(\mu_0,\sigma),y_k(\mu_0, \sigma))$. We claim that a sequence
$\zeta=\left\{(x_k(\mu_0, \sigma),y_k(\mu_0, \sigma))\right\}$ is a
$\delta$-pseudotrajectory for the dynamical system $S_{\mu^*}(x_0,
y_0)$, with $\delta=\sqrt{2}|\mu^*-\mu_0|+\mathcal{K}\sigma$,
$\mathcal{K} \in \rr$. To obtain this result we compute the Euclidean
distance:
\begin{equation}
\label{delta_dist}
  d(S_{\mu^*}(x_{k}(\mu_0,\sigma),y_{k}(\mu_0,\sigma)),
  (x_{k+1}(\mu_0,\sigma),y_{k+1}(\mu_0,\sigma)))
\end{equation}
For the sake of simplicity $(\bar{x}_{k+1},\bar{y}_{k+1})$ are the
observations, i.e.  Gaussian random variables with mean $x_{k+1}$
($y_{k+1}$, respectively), and standard deviation $\sigma$, as in
Sec.~\ref{sec:OD_stmap}, and
$S_{\mu^*}(\bar{x}_{k},\bar{y}_{k})=(\tilde{x}_{k+1},\tilde{y}_{k+1})$.
Using these notations, (\ref{delta_dist}) turns into
\begin{equation*}
  d(S_{\mu*}(\bar{x}_{k},\bar{y}_{k}),
  (\bar{x}_{k+1},\bar{y}_{k+1})) =
  \sqrt{(\tilde{x}_{k+1}-\bar{x}_{k+1})^2+(\tilde{y}_{k+1}-\bar{y}_{k+1})^2}
\end{equation*}
We compute separately the two differences.
\begin{eqnarray}
\label{eq:diff_y}
  |\tilde{y}_{k+1}-\bar{y}_{k+1}|&=&|\bar{y}_{k+1}-\mu^*\sin{\bar{x}_k}-y_{k+1}-\mathcal{N}(0,\sigma^2)|\nonumber\\
  &=&|\mathcal{N}(0,2\sigma^2)-\mu^*\sin{x_k}\cos({\mathcal{N}(0,\sigma^2)})+\mu^*\sin({\mathcal{N}(0,\sigma^2)})\cos{x_k}+\mu_0\sin{x_k}|\nonumber\\
&<&\mathcal{N}(0,2\sigma^2)+|\mu_0-\mu^*|
\end{eqnarray}
To allow the last estimate, we need to solve a technical problem: the
Shadowing Lemma uses a uniform norm, that is the maximum of the
distance between the $\delta$-pseudotrajectory and the
$\varepsilon$-shadowing. On the contrary, the natural norm for the
residuals of the fit is the Euclidean norm with the square root of the
sum of the squares. However, since the number of residuals is not only
finite but sharply limited by the numerical phenomena discussed in
Section~\ref{sec:numerical}, in a given experiment we can just take
the maximum absolute value of the residuals and it is going to be
${\cal K}\sigma$, with ${\cal K}$ a number which in practice cannot be
too large, e.g., in our experiment ${\cal K}=5.9$.

Then we can approximate the quantities ${\cal O}(\sigma)$ and smaller,
e.g., $\cos({\mathcal{N}(0,\sigma^2)}) \sim 1$ and
$\sin({\mathcal{N}(0,\sigma^2)}) \sim 0$.
The $x$ coordinate gives a similar result:
\[
|\tilde{x}_{k+1}-\bar{x}_{k+1}|=|\bar{x}_{k+1}+\tilde{y}_{k+1}-x_k-y_{k+1}
-\mathcal{N}(0,\sigma^2)|
\]
\begin{equation}
\label{eq:diff_x}
< \mathcal{N}(0,\sigma^2)+\mathcal{N}(0,2\sigma^2)+|\mu_0-\mu^*|=
\mathcal{N}(0,3\sigma^2)+|\mu_0-\mu^*|
\end{equation}
Putting together (\ref{eq:diff_y}) and (\ref{eq:diff_x}) we obtain
\begin{gather}
  \begin{aligned}
    \sqrt{(\tilde{x}_{k+1}-\bar{x}_{k+1})^2+(\tilde{y}_{k+1}-\bar{y}_{k+1})^2}&< \\
   < \sqrt{(\mathcal{N}(0,3\sigma^2)+|\mu_0-\mu^*|)^2+(\mathcal{N}(0,2\sigma^2)+|\mu_0-\mu^*|)^2}&<\\
   < \sqrt{2}|\mu_0-\mu^*| + \sqrt{(\mathcal{N}(0,3\sigma^2))^2+(\mathcal{N}(0,2\sigma^2))^2}< \sqrt{2}|\mu_0-\mu^*| +\mathcal{K}\sigma\
  \end{aligned}\label{eq:deltaps}
\end{gather}
with $\mathcal{K} \in \rr$.

Therefore the sequence generated by the observations is a
$\delta$-pseudotrajectory for the dynamical system $S_{\mu^*}$ with
$\delta=\sqrt{2}|\mu_0-\mu^*|+\mathcal{K}\sigma$.

Figure~\ref{fig:deltapseudo_matlab} is an example of observations as a
$\delta$-pseudotrajectory. The observations are built with
$\sigma=10^{-3}$ and $\mu_0=0.5$, and the dynamical system is
$S_{\mu^*}$, with $|\mu^*-\mu| = 10^{-1}$; the circles have radius $\delta$.
\begin{figure}[h!]
  \figfig{8cm}{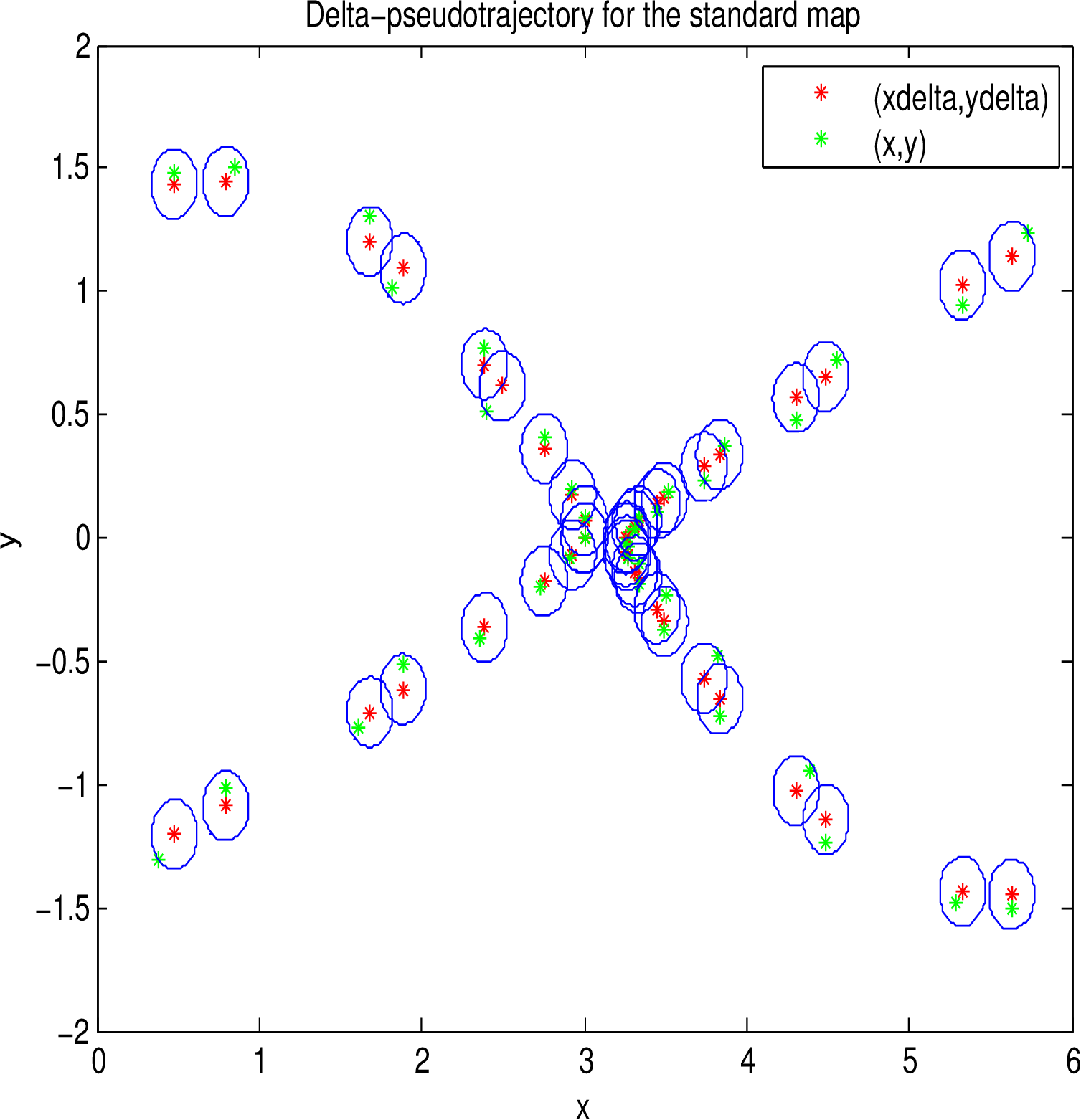}{An example of a
    $\delta$-pseudotrajectory. Initial conditions are $x_0=3$,
    $y_0=0$, $\mu_0=0.5$. $\delta \mu=10^{-1}$, and $\sigma=10^{-3}$.}
\end{figure}

The solution of the least squares fit (to the observations from $-n$
to $n$), obtained by convergent differential corrections, is a finite
$\varepsilon$-shadowing, valid for the iterations from $-n$ to $n$.

We choose a value $\varepsilon>0$, that is a boundary on the maximum
of the residuals. Then we choose $\sigma<\varepsilon/{\cal K}$ and we set up
the observations process. Next, we create a first guess: a new orbit
obtained with a small change of the initial conditions and of the
dynamical parameter $\mu$:
$\{(x_k(\mu_g),y_k(\mu_g))\}=S_{\mu_g}(x_g,y_g)$, with $x_g=x_0+dx$,
$y_g=y_0+dy$, and $\mu_g=\mu_0+d\mu$. Then we apply the differential
corrections to the orbit. If the iterations converge, that is the last
correction is very small, the maximum of the norm of the residuals is
less than $\varepsilon$ (because the individual residuals are less
than $3\sigma$).

At convergence, we obtain an initial condition
$(x^*,y^*)$ and a value of the dynamical parameter $\mu^*$, such that
$(x^*,y^*)$ is the $(\varepsilon,S_{\mu^*})$-shadowing for the
$\delta$-pseudotrajectory with
$\delta=\sqrt{2}|\mu^*-\mu_0|+\mathcal{K}\sigma$, for all the points
used in the fit.

The most important requirement is the convergence of the differential
corrections, otherwise we cannot obtain the
$\varepsilon$-shadowing. This is far from trivial, because the chaotic
divergence of the orbits introduces enormous nonlinear effects, for
which the linearized approach of differential corrections may
  fail. To guarantee convergence, first we select the initial
conditions $x_0,y_0$ to be at the center of the observed interval,
otherwise the initial conditions would be essentially undetermined
along the stable manifold of the initial conditions. Second, we use a
\textit{progressive solution} approach, namely, given the solution
with $2n+1$ observations with indexes between $-n$ and $n$, we use the
convergent solution $x_0^*, y_0^*, \mu^*$ for $n$ as first guess for
the solution with $2n+3$ observations (between $-n-1$ and $n+1$).
Then the initial guess is actually used only for the solution with $3$
observations, for which the nonlinearity is negligible.  Still the
convergence of the differential corrections depends critically upon
the number $n$ of iterations of the map, as explained in
Section~\ref{sec:numerical}.

\section{Numerical results}
\label{sec:numerical}

The experiment was performed with the initial conditions at $x_{0}=3$
and $y_{0}=0$, and the value of the dynamical parameter
$\mu_0=0.5$. The dynamical context for this orbit can be appreciated
from Figure~\ref{smap_zoom}, showing that the initial conditions
are indeed in a portion of the initial conditions space
containing mostly chaotic orbits. For the observation noise we have
used a standard deviation $\sigma=10^{-10}$.

\subsection{Computability horizon}
\label{sec:horizon}
\begin{figure}[h!]
  \figfig{10cm}{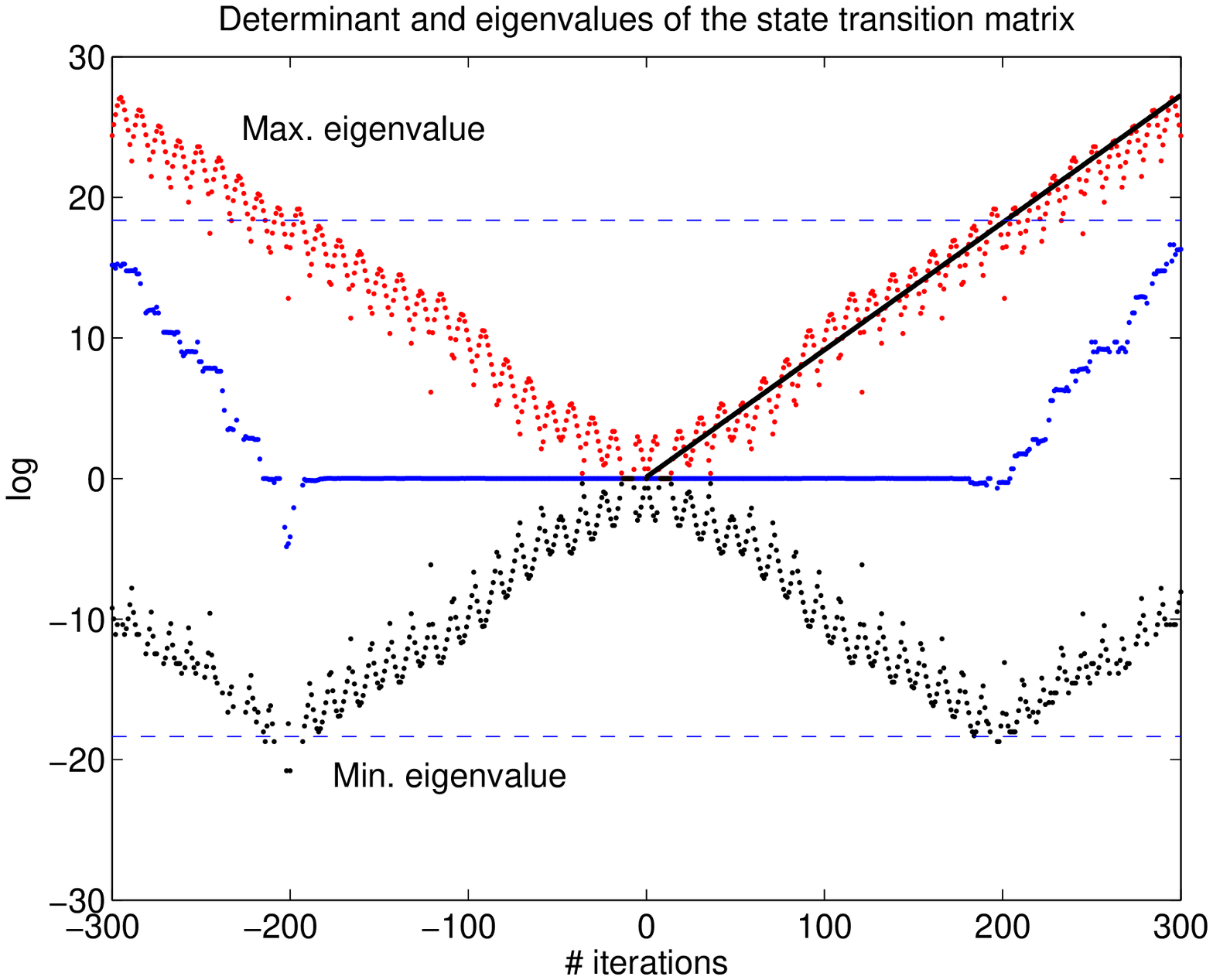}{The eigenvalues and the
    determinant of the state transition matrix in a semilogarithmic
    scale, as a function of the number of iterations. Also shown is
    the linear fit to the large eigenvalue based on the first $180$
    iteration, with slope $+0.091$.  The computation is in double
    precision and the number of iterations of the standard map $n$ is
    $300$ with the map and $300$ with its inverse. The determinant of
    the state transition matrix would be $1$, for all $n$, in an exact
    computation. The numerical instability occurs when the eigenvalues
    reach the critical values $\sqrt{\varepsilon_d},
    \sqrt{1/\varepsilon_d}$ marked by the dotted lines.}
\end{figure}

\begin{figure}[h!]
  \figfig{10cm}{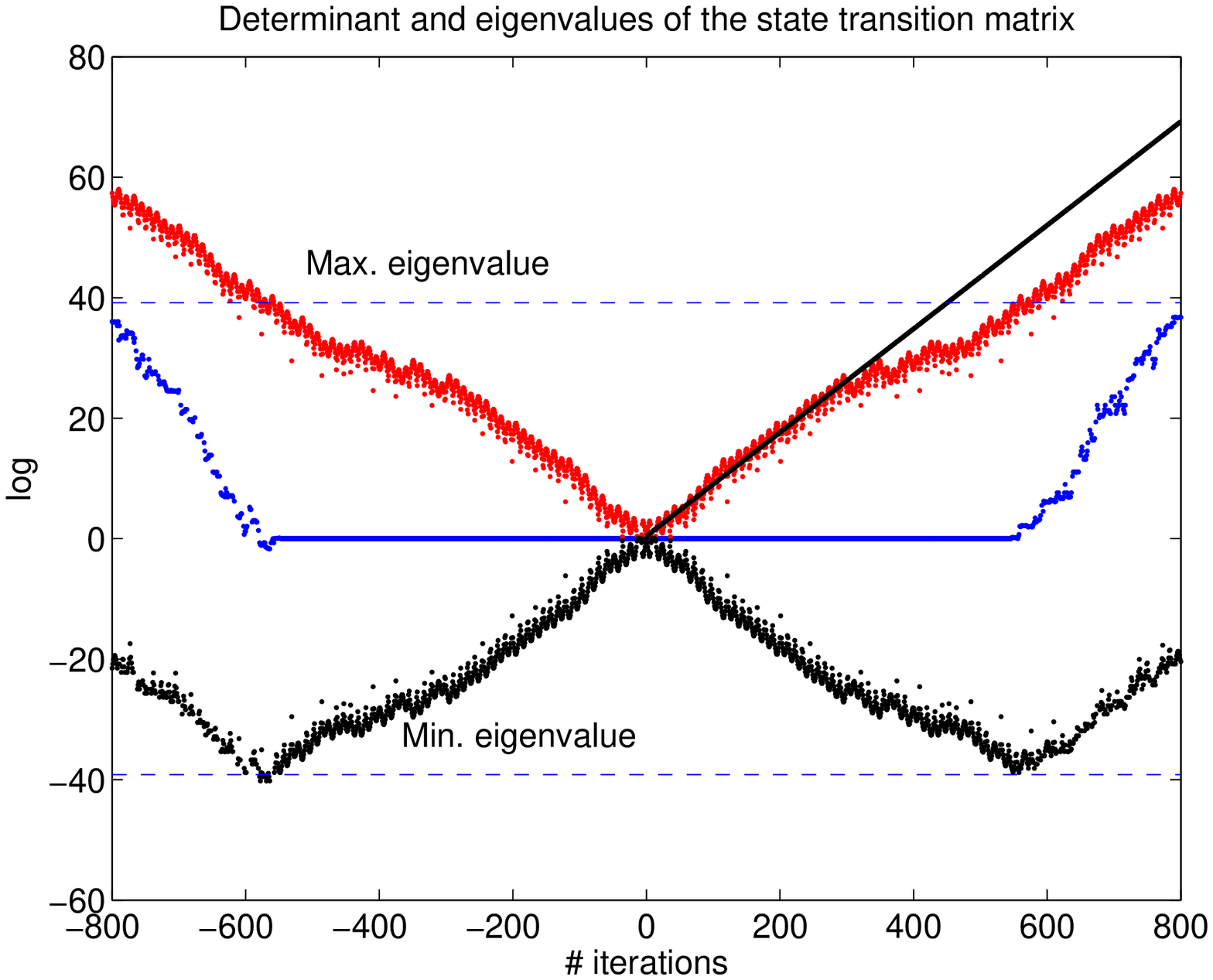}{The eigenvalues and the determinant
    of the state transition matrix in a semilogarithmic scale, as a
    function of the number of iterations. Also shown is the
    linear fit to the large eigenvalue based on the first
    $300$ iteration, with slope $+0.086$.  The computation is in
    quadruple precision and the number of iterations of the standard
    map $n$ is $800$ with the map and $800$ with its inverse. The
    numerical instability occurs when the eigenvalues reach the
    critical values $\sqrt{\varepsilon_q}, \sqrt{1/\varepsilon_q}$
    marked by the dotted lines.}
\end{figure}

Figures~\ref{fig:det_eigenvalues_double} and \ref{fig:det_eigenvalues}
show the absolute value of the eigenvalues of the state transition
matrix forward and backward. The product of two eigenvalues should be
$1$ in exact arithmetic. When the condition number of the matrix
becomes larger than the inverse of the machine rounding off error, the
computation of the matrix becomes numerically impossible, and the
computed value of the determinant is far from $1$.

In Figure~\ref{fig:det_eigenvalues_double} the computations are
performed in the standard double precision, that is with a mantissa of
$52$ binary digits and a round off relative error of
$\varepsilon_d=2^{-53}=1.1\times 10^{-16}$. We observe a numerical
instability after $\simeq 180$ iterations: the determinant
deviates from the exact value of $1$ and the small
eigenvalue starts increasing; the large eigenvalue keeps increasing,
but there is a slight change of slope. Then we fit the slope
of the large eigenvalue curve for the first $180$ iterations, and get
a Lyapounov indicator $+0.091$: it approximates the maximum Lyapounov
exponent $\chi$ for the orbit to which our differential corrections
converge\footnote{There is no way to rigorously compute the Lyapounov
  exponents: in practice \textit{Lyapounov indicators} extracted from
  finite propagations are used to assess, but not rigorously prove,
  the chaotic nature of the orbits. Note that it is a numerically well
  documented phenomenon that the indicators are not constant, but
  actually depend upon the time interval over which they are computed,
  although in most cases these changes are not very large and the
  conclusion that an orbit is chaotic is reliable.}.

The Lyapounov time is $T_L=1/\chi$, in this example $T_L\simeq 11$. To
reach a ratio of eigenvalues of the state transition matrix of
$1/\varepsilon_d$ we need a number of Lyapounov times
$\log(1/\sqrt{\varepsilon_d})$, in this case $\simeq 18.4\,T_L\simeq
202$ iterations of the map.  At about this number of iterations the
maximum and minimum eigenvalues of $A_n$ are so widely apart in size
that a bad conditioning horizon is reached, and the computation of the
state transition matrix becomes numerically inaccurate.  Hence near
$\pm 18.4\, T_L$ we observe the numerical instability in the
computation of the determinant and of the eigenvalues of the state
transition matrix.

Figure~\ref{fig:det_eigenvalues} shows the same computations, with the
same initial condition, but in quadruple precision, with a $112$ bit
mantissa and $\varepsilon_q=2^{-113}=9.6\times 10^{-35}$. The change
of slope in the eigenvalues curves occurs after $\simeq 300$
iterations, while a full blown numerical instability occurs after
$\simeq 550$ iterations. The fit to the large eigenvalue for the first
$300$ iterations gives a Lyapounov indicator $+0.086$, not very
different from the one obtained in double precision. Thus we would
expect the numerical instability to occur after
$\log(1/\sqrt{\varepsilon_q})\,T_L\simeq 39.2\, T_L=455$
iterations. It appears that the rate of divergence decreases after
$300$ iterations, as shown by the change in slope, allowing to
maintain at least the determinant near the exact value for about $100$
more iterations beyond the value predicted above.

The \textit{computability horizon} represents the maximum number of
iterations we can reach, before the computation becomes numerically
unstable. The computability horizon strongly depends on the
chaoticity of the system: more chaos, that is larger $\chi$, more
instability; but also upon the precision of the computations.

Thus, in the following we perform the numerical experiments in
quadruple precision, in order to mitigate the problem of the numerical
instability. We compute $500$ iterations forwards and backwards, but
we use only the first $300$ iterations for the linear fits, to avoid
the possibility that changes in slope, such as the ones apparent in
Figure~\ref{fig:det_eigenvalues}, contaminate our experimental
results.

The compatibly horizon is a hard limit in that it is not
practically possible to increase the number of iterations to a much
higher value. E.g., to push the horizon by a factor $10$
above the value for double precision, we would need computations
performed with real numbers represented with $800$
bytes\footnote{Software to perform arithmetic computations with an
  arbitrary number of digits is available, but the algorithms are too
  slow to be used even for our simple example.}.

The conclusion is that the practical problem of chaotic orbit
determination is meaningful only for a finite number of iterations,
and the accuracy of the results can be tested only within the boundary
of the computability horizon. 

\subsection{Chaotic case}
\label{sec:chaos}

Figure~\ref{fig:chaos_unc_3par} shows the results in quadruple
precision of the full 3-parameter fit: the 3 parameters are the
initial conditions and the dynamical parameter $\mu$. The
determination of $\mu$ is indeed not possible without simultaneous
determination of the initial conditions.

Even in quadruple precision we find a maximum value of $n$ beyond
which the iterative solution of the nonlinear least squares problem is
divergent. This maximum turns out to be $599$ in this experiment: it
is close to what we have called the computability horizon, that
is this limitation is due to the difficulty of computing the state
transition matrix when the condition number is too large.

\begin{figure}[h!]
  \figfig{10cm}{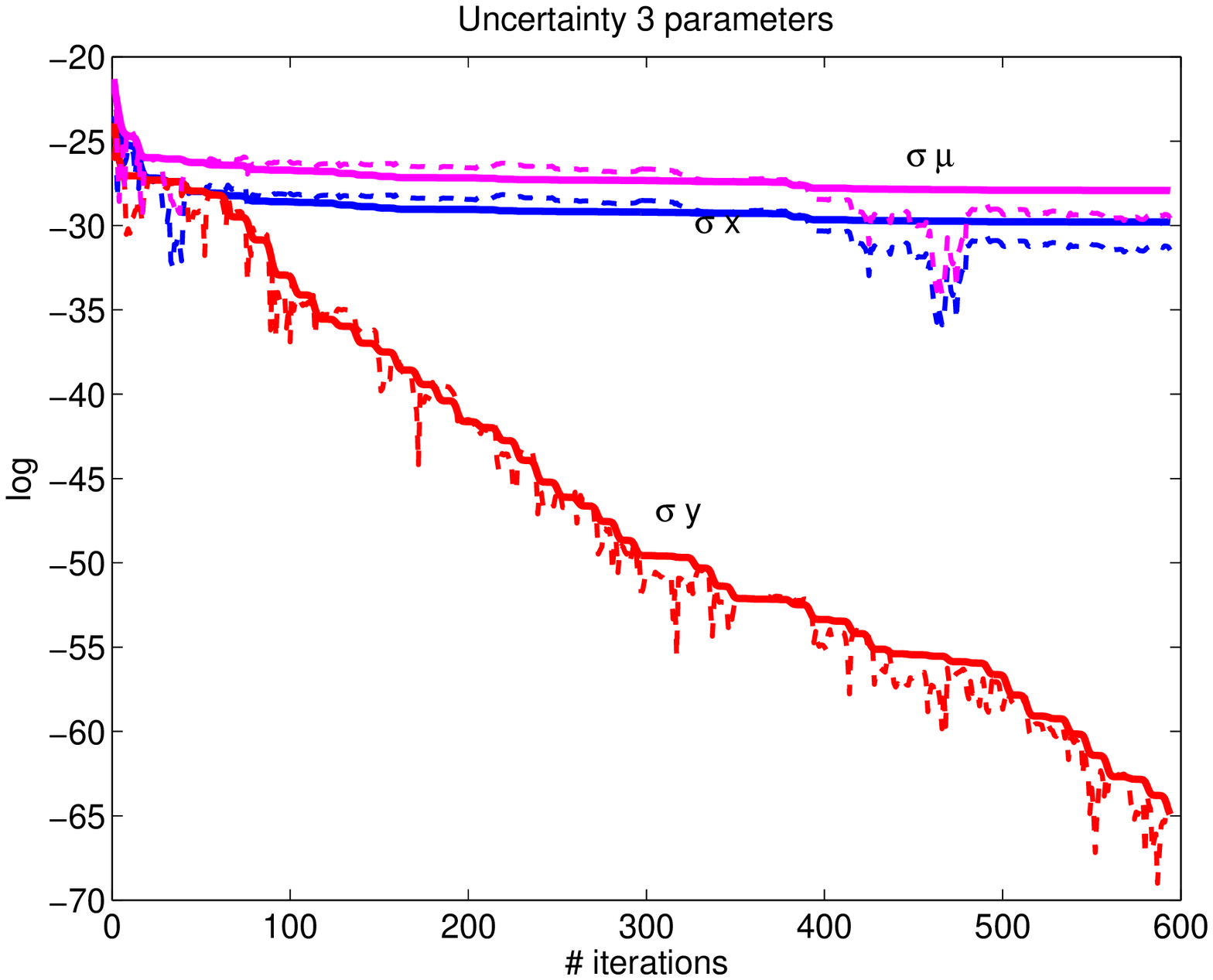}{Standard deviation of the solutions
    for the initial conditions and for the dynamical parameter $\mu$
    (continuous lines), and actual error (nominal solution of
      the fit minus real value used in the simulation, dashed lines),
    as a function of the number of iterations.}
\end{figure}

The curves in Fig.~\ref{fig:chaos_unc_3par} represent both the formal
standard deviation and the actual error of the solutions of the least
squares fit, as a function of $n$ in a semilogarithmic plot. Both the
formal standard deviation and the actual error of $\mu$ do not
decrease exponentially. Indeed, in Fig.~\ref{fig:chaos_unc_3par_fit}
we have the same behavior of the curves that we have already seen in
Fig.~\ref{fig:chaos_unc_3par}, but in a log-log plot, in
which a constant slope $a$ would imply a power law proportional to
$n^a$. The slopes of the lines that fit the uncertainties are:
$-0.675$ for the dynamical parameter $\mu$, $-0.833$ and $-12.030$ for
the initial conditions $x$ and $y$, respectively. This plot in
logarithmic scale shows that the uncertainty for $\mu$ and $x$ does
not decrease exponentially.  It is also apparent that one of the
initial conditions ($y$) is better determined than the other one
($x$), with an improvement as a function of $n$ which could be
exponential. This is a property of the specific initial condition we
have used, for other choices we can get three parameters determined
with comparable accuracy, none of them with exponential
improvement\footnote{This depends upon the orientation of the stable
  and unstable directions at the initial condition.}.

\begin{figure}[h!]
  \figfig{10cm}{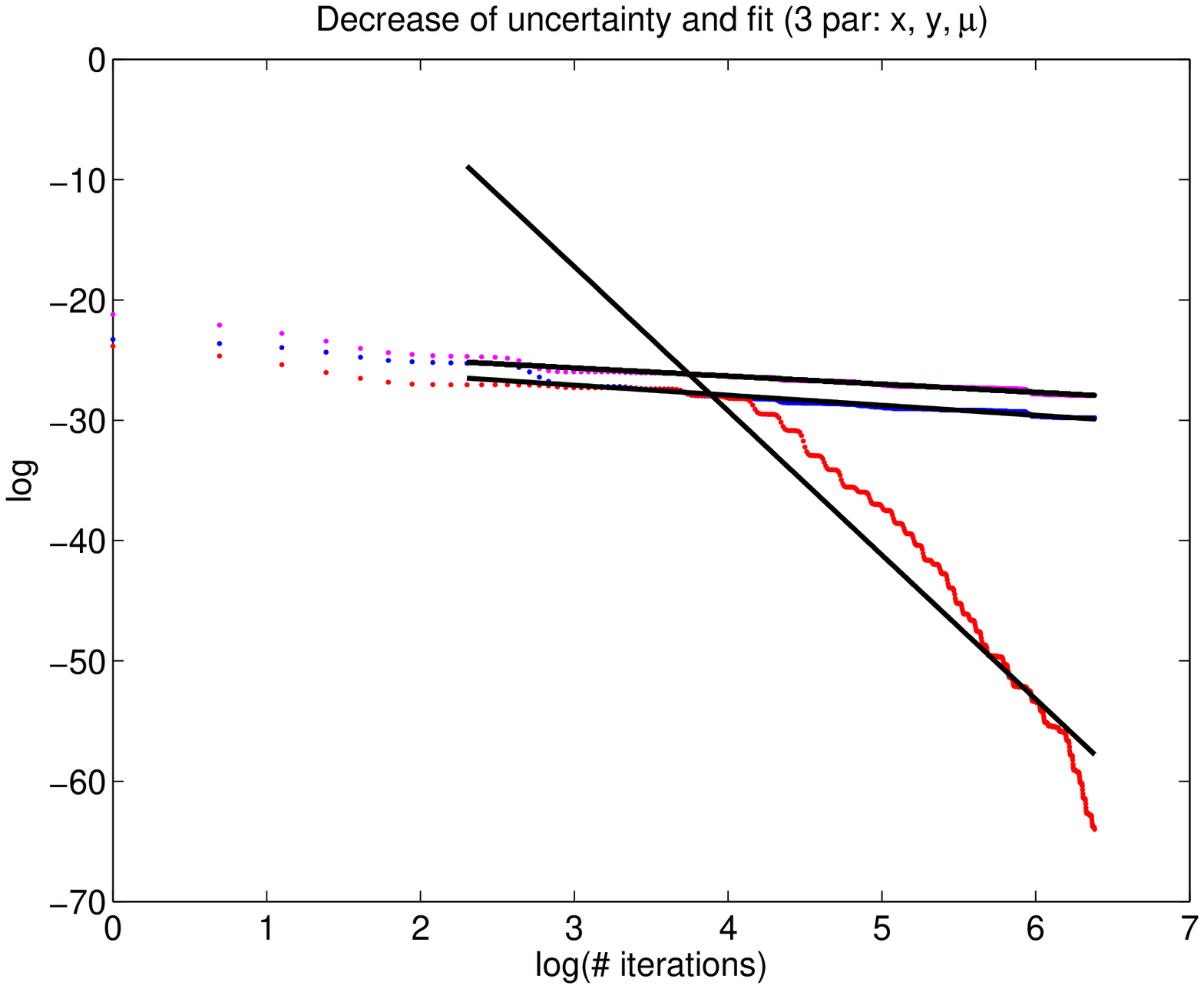}{Uncertainty of the solution of the
    least squares fit for the initial conditions and for the dynamical
    parameter $\mu$ in a logarithmic scale. }
\end{figure}

Figure~\ref{fig:chaos_unc_2par} shows the results for the standard
deviation and the actual error when solving only for the initial
condition. The 2x2 portion of the normal matrix which refers only to
the initial conditions is not badly conditioned.  Also as a result of
this, we are able to get convergence of the differential corrections
up to $\pm 742$ iterations, which is even beyond the numerical
stability boundary. If the fit is done by using only up to $300$
iterates, to avoid the apparent slope change, the slopes shown in this
Figure are $-0.084$ for $x$ and $-0.083$ for $y$; note that the
Lyapounov indicator for the same interval is $+0.086$. Thus
exponentially improving determination of the initial conditions only
is possible, and the exponent appears to be very close to the opposite
of the Lyapounov exponent.

\begin{figure}[h!]
  \figfig{10cm}{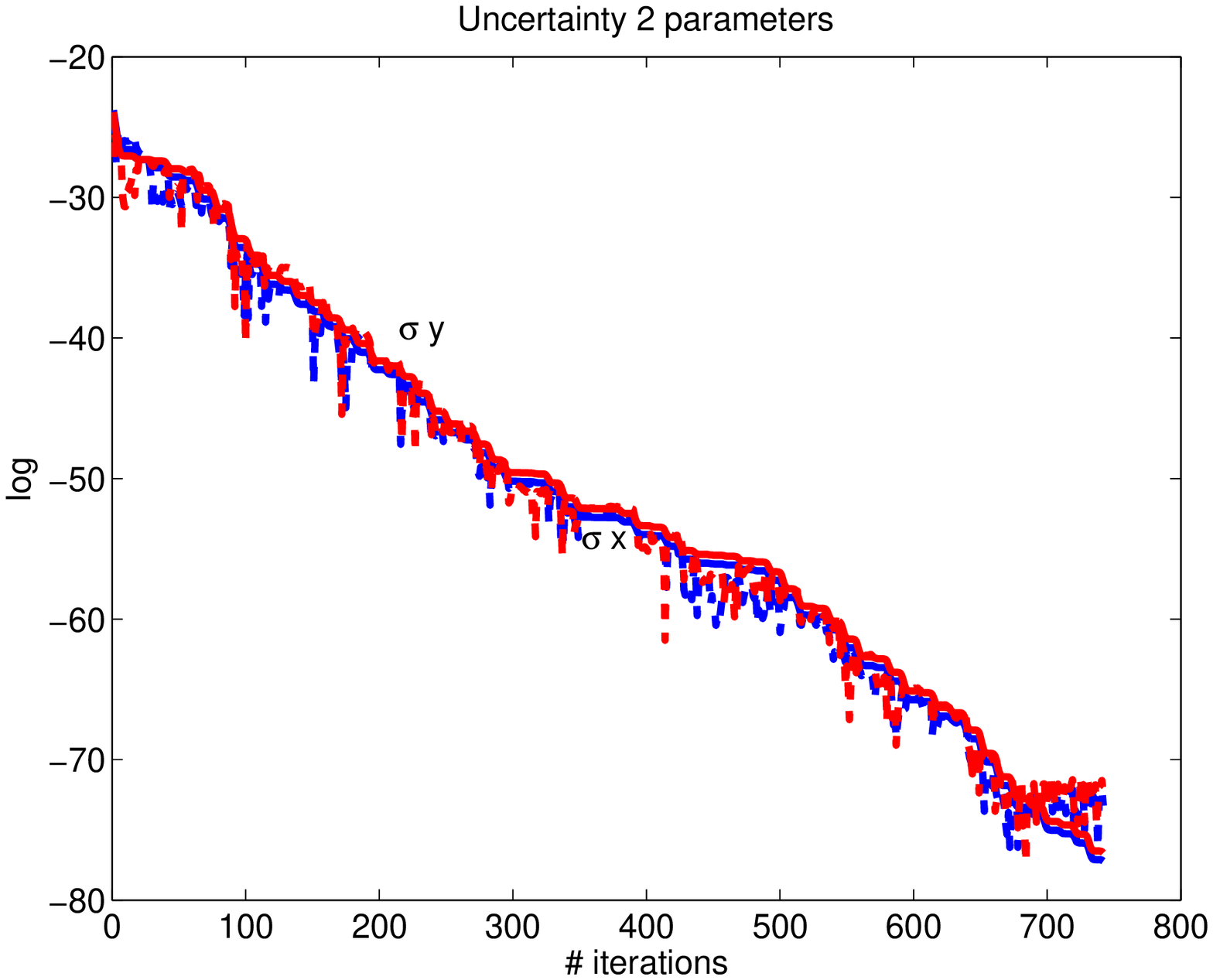}{Standard deviation of the solutions
    for the initial conditions (continuous lines), true errors for the
    same 2 parameters (dashed lines).}
\end{figure}

\subsection{Ordered case}
\label{sec:ord}

An ordered case can be obtained with a change of the initial
conditions. For the numerical experiments we have chosen $x_0=2$,
$y_0=0$ and $\mu_0=0.5$. In the ordered case we have not the problem
of the computability horizon and the Lyapounov exponent is very
small: actually, it could be zero if we are on a Moser invariant curve.
Thus we have computed $5000$ iterations. 

\begin{figure}[ht] 
  \begin{minipage}[b]{0.6\linewidth}
    \centering
    \includegraphics[width=.8\linewidth]{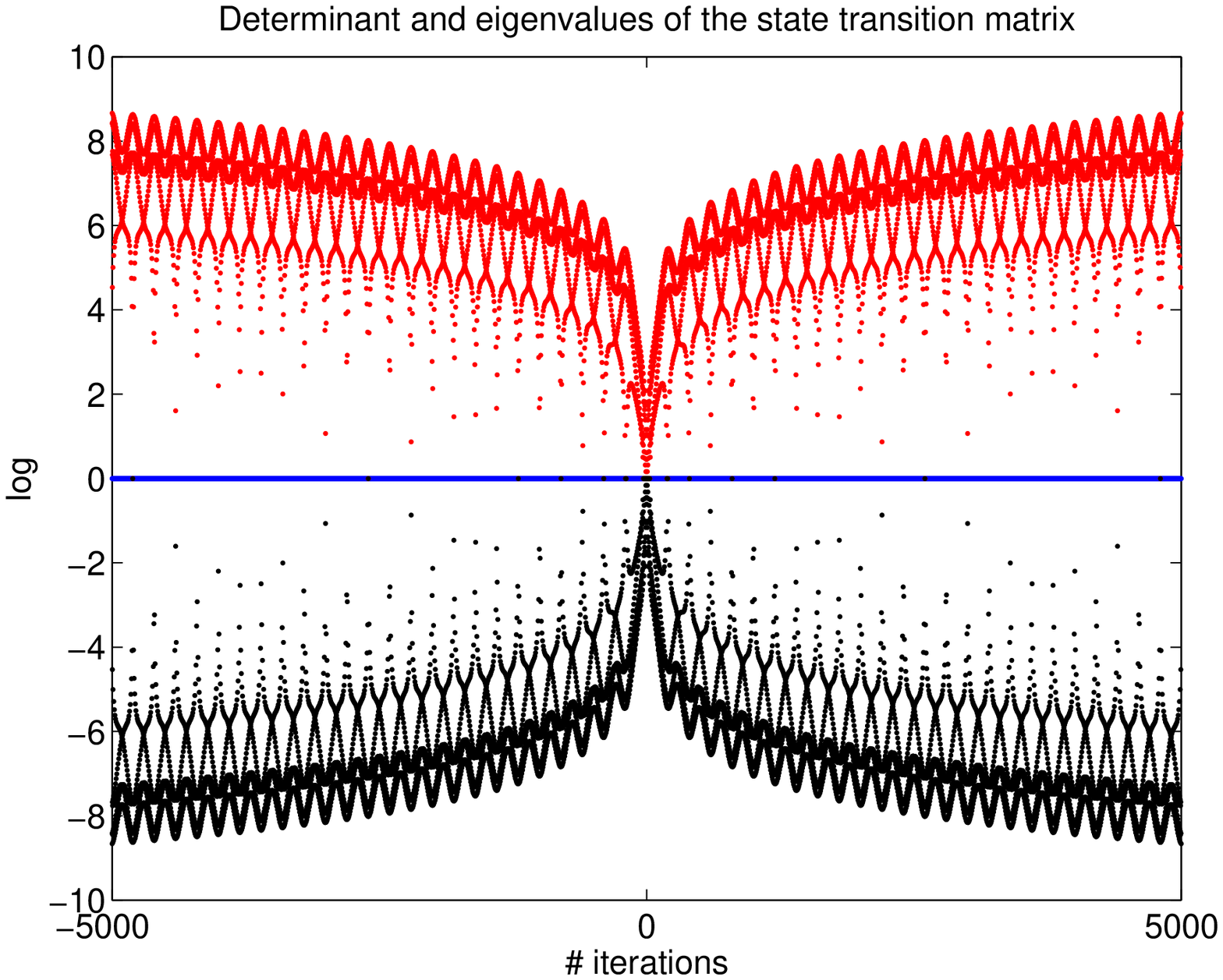} 
    \vspace{4ex}
  \end{minipage}
  \begin{minipage}[b]{0.6\linewidth}
    \centering
    \includegraphics[width=.8\linewidth]{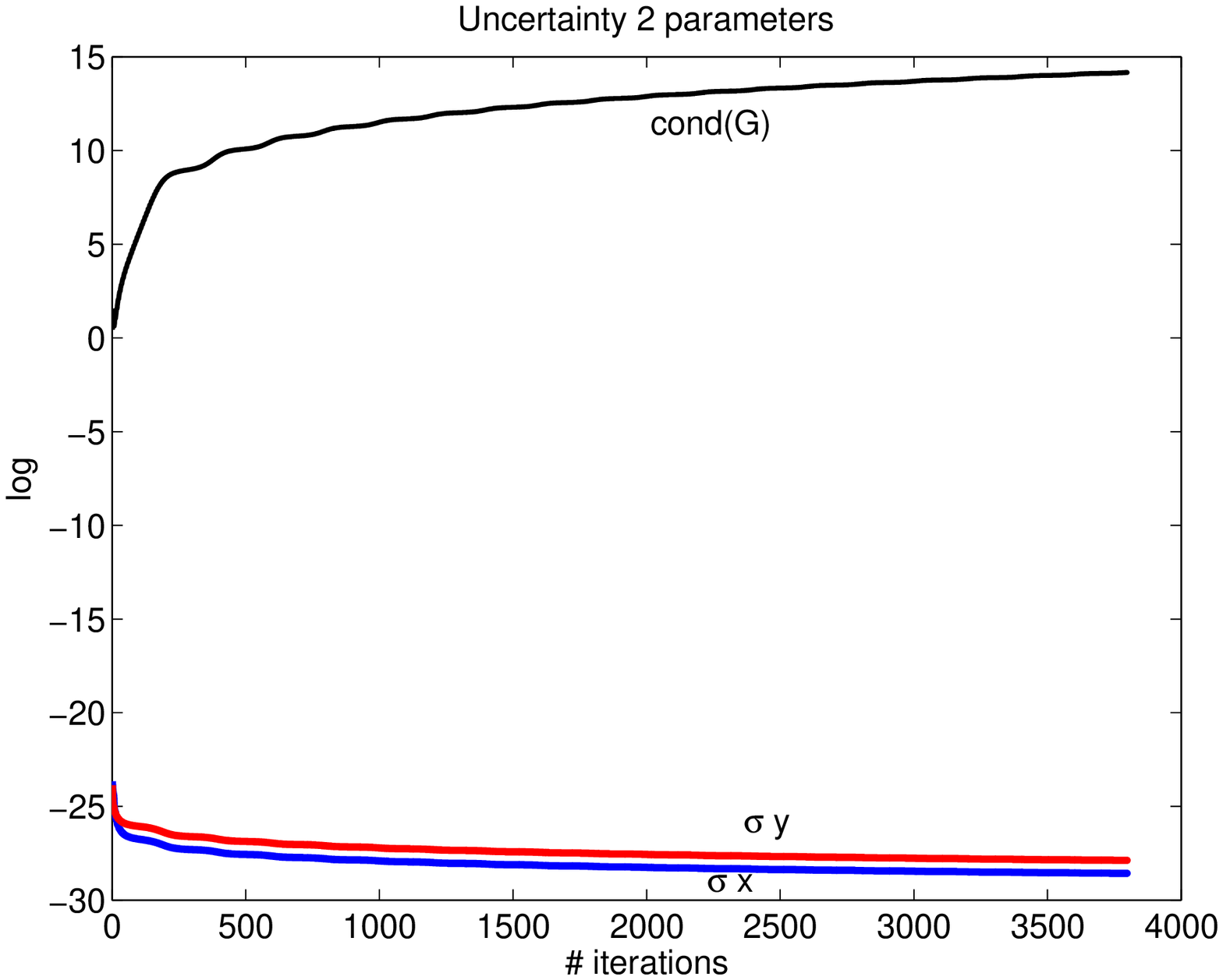} 
    \vspace{4ex}
  \end{minipage} 
  \begin{minipage}[b]{0.6\linewidth}
    \centering
    \includegraphics[width=.8\linewidth]{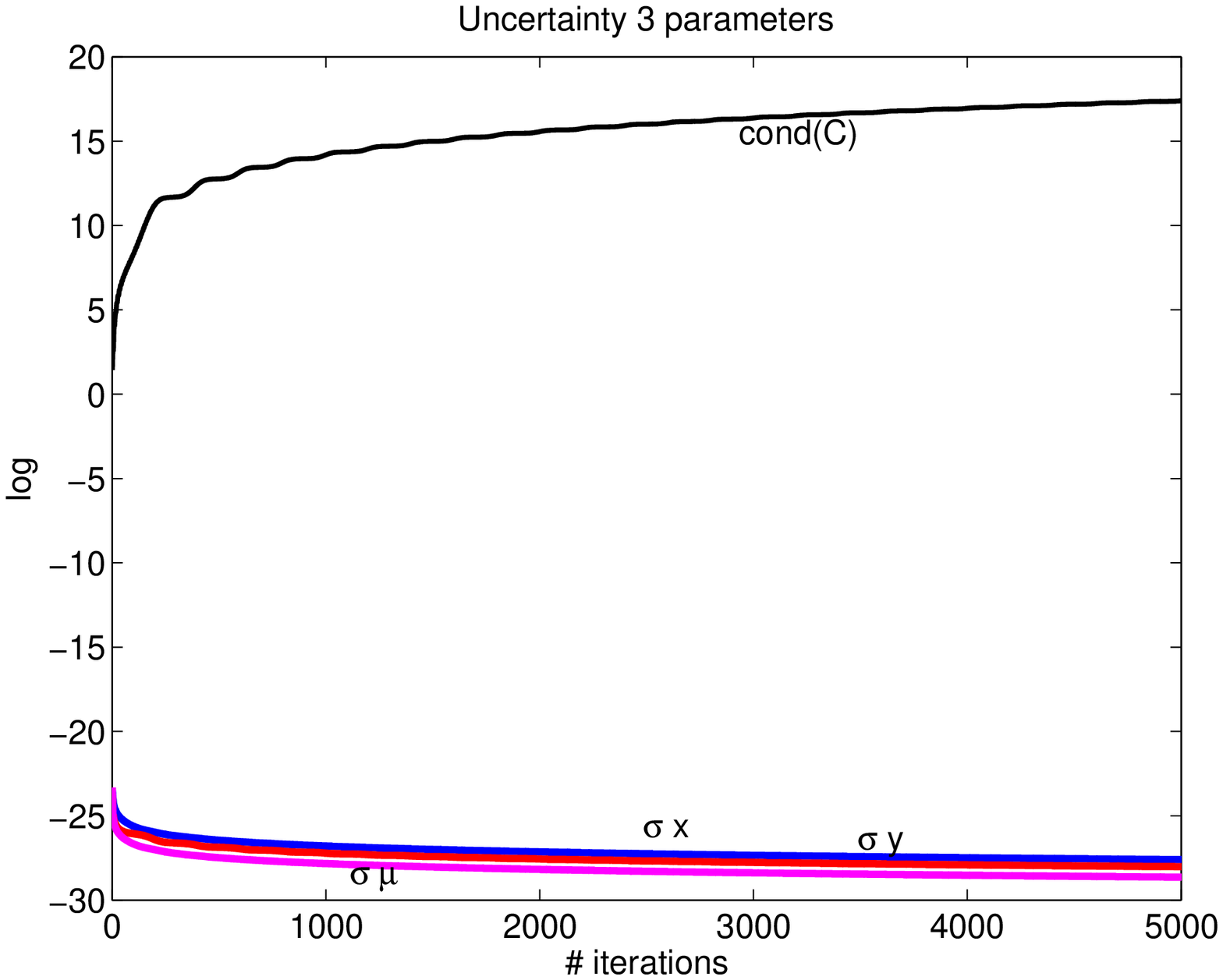} 
    \vspace{4ex}
  \end{minipage}
  \begin{minipage}[b]{0.6\linewidth}
    \centering
    \includegraphics[width=.8\linewidth]{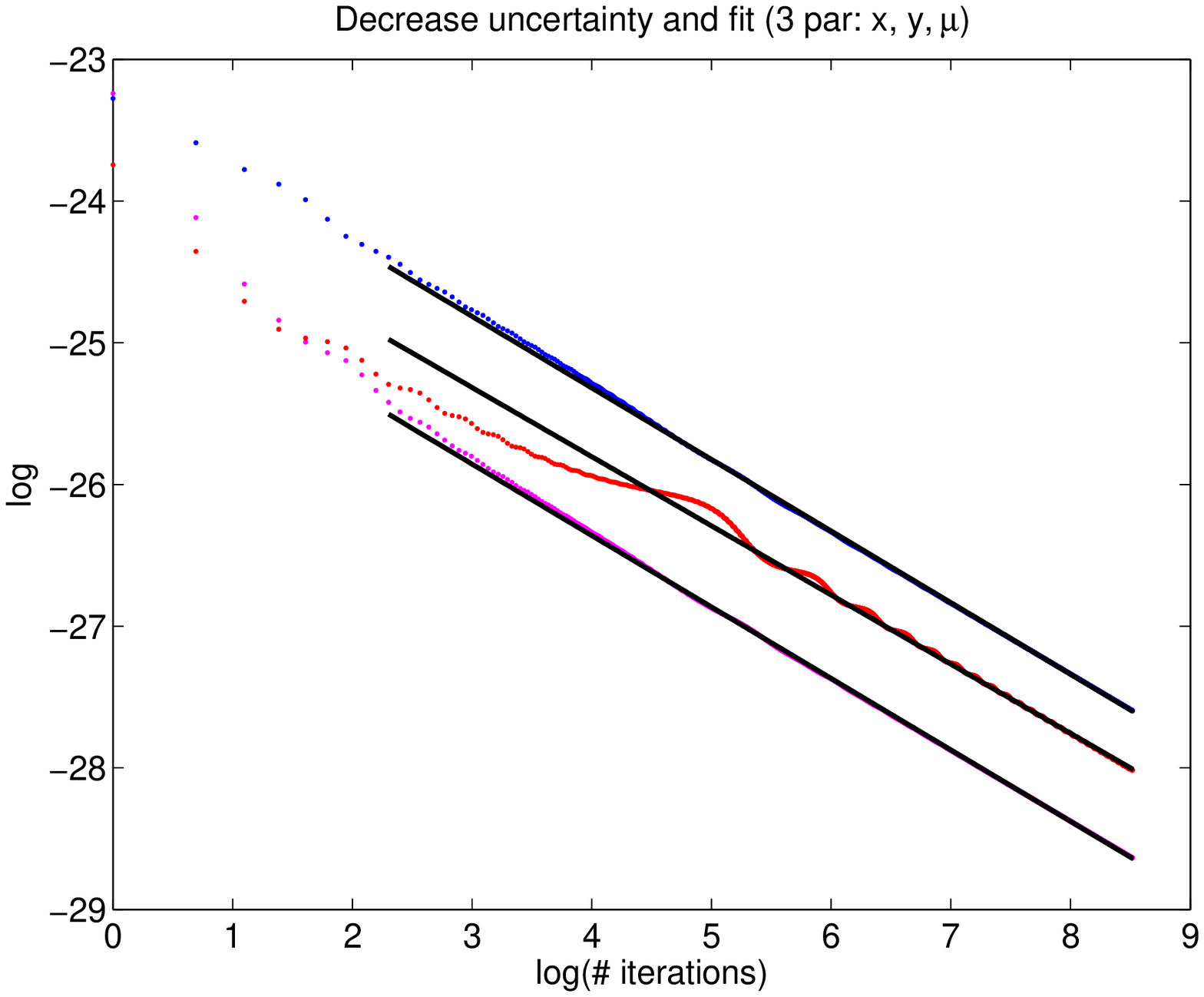} 
    \vspace{4ex}
  \end{minipage} 
  \caption{Top left: eigenvalues of the state transition matrices, for
    the chosen regular initial conditions and for $\pm 5000$
    iterations. Top right: solutions for the initial condition only,
    from the top condition number of the normal matrix,
    standard deviation of $y$ and for $x$. Bottom left: solutions for
    three parameters, from the top condition number, standard
    deviation of $x$, of $y$, of $\mu$. Bottom right: log-log plot of
    the 3 standard deviations, with very similar slopes.}
\label{fig:ordered}
\end{figure}

Figure~\ref{fig:ordered} gives a summary of our numerical
experiment in the ordered case. The Lyapounov indicator is very small
($\simeq 10^{-4}$), and can be made even smaller by continuing the
experiment for larger values of $n$. As a consequence, the state
transition matrix is not badly conditioned, and the computability
horizon is much beyond the number of iterations we have used (if it
exists at all). Thus the lack of chaoticity implies the practical
absence of the computability horizon, and we can determine all the
parameters with very good accuracy, even if we are not in exact
arithmetic.  The values of the slopes of the fit to the uncertainty
are $-0.504$ for $\mu$, $-0.504$ and $-0.488$ for $x$ and $y$
respectively, the corresponding regression lines are shown in the
log-log plot on the bottom right. As it is clear by comparing the top
right and the bottom left plot, the standard deviation for the
solution with only 2 parameters have very much the same behavior,
indeed in a log-log plot (not shown) we can get slopes $-0.511$ for
$x$, $-0.481$ for $y$.

All these power laws are close to the inverse square root of the
number of iterations, namely the same rule as the standard deviation
in the computation of a mean. We do not have a formal proof of this,
but we conjecture that for an orbit on a Moser invariant curve (for
which the Lyapounov exponents are exactly zero) the standard
deviations for all the parameters decrease as $1/\sqrt{n}$.

\section{Conclusions}
\label{sec:conclusions}

We have understood the concept of \textit{computability horizon} as a
consequence of numerical instability in the computation of the state
transition matrices, thus providing a comparatively simple empirical
formula to approximately predict the horizon. This is a practical
limitation which applies to any attempt at orbit determination of
chaotic orbits.

We have used numerical experiments in quadruple precision, but
nevertheless limited by the computability horizon to few hundreds of
iterations for the very chaotic orbit used in our test. From these we
have found the following three empirical facts:
\begin{enumerate}
\item If only initial conditions are determined for a chaotic orbit,
  the uncertainty can decrease exponentially with the number $n$ of
  iterations of the map, and the exponent of this decrease is close to
  a Lyapounov exponent.
\item If a dynamical parameter $\mu$ is determined together with the
  initial conditions of a chaotic orbit, the decrease in uncertainty
  is polynomial in $n$ for $\mu$ and for at least one of the initial
  coordinates.
\item If the initial conditions belong to an apparently ordered orbit,
  that is such that there is no evidence of a positive Lyapounov
  exponent, it is possible to determine simultaneously $\mu$ and the
  initial conditions with uncertainty decreasing polynomially with
  $n$. The case in which only the initial conditions are determined
  gives the same result. Moreover, all the power laws $n^a$ for these
  uncertainties appear numerically to have $a\simeq -1/2$.
\end{enumerate}

\subsection{Discussion on the Wisdom hypothesis}

\label{sec:discwisd}

The statement by Wisdom, as a practical rule for concrete orbit
determination, appears to be first limited by the computability
horizon. Second, the actual decrease of the uncertainty, going as far
as it can be done numerically, is not exponential, but polynomial, as
$n^a$, with $a$ negative and rather small, although we have found that
the value of $a$ depends upon the initial conditions\footnote{We are
  showing figures and giving data only for one initial condition, but
  of course we have run many tests.}. Note that the orbit
determinations in which the only parameters to be solved are the 2
initial coordinates show an exponential decrease as $\exp(-\alpha\,
n)$, where $\alpha$ appears to be close to the Lyapounov exponent
$\chi$, but the strong correlations appearing when 3 parameters are
solved degrade the result in a very substantial way.

This needs to be compared to the regular case, shown in
Figure~\ref{fig:ordered}, where the standard deviations for each of
the 3 fit parameters decrease approximately according to an
$1/\sqrt{n}$ law, as prescribed by the standard rule for the estimate
of the mean with errors having a normal distribution. Indeed it is
possible that the determination of $\mu$ for some chaotic cases,
including the example shown in Figure~\ref{fig:chaos_unc_3par_fit},
decreases faster than for an ordered case, but the decrease is anyway
polynomial, proportional to $n^a$ with some different negative $a$,
thus the difference is not very large, given the tight constraint on
the maximum possible value of $n$.

\subsection{Examples from Impact Monitoring}
\label{sec:impacts}

One feature of our results is that adding a dynamical parameter to the
list of parameters to be determined results in degradation in the
normal matrix, thus in much slower decrease of the uncertainties as
the number of observations grows. The problems of orbit determination
for NEA undergoing several close approaches to the Earth (or other
planets) is more complex than our simple model, but we have found that
the phenomenon described above does occur in a remarkably similar way.

\begin{figure}[h!]
  \figfig{10cm}{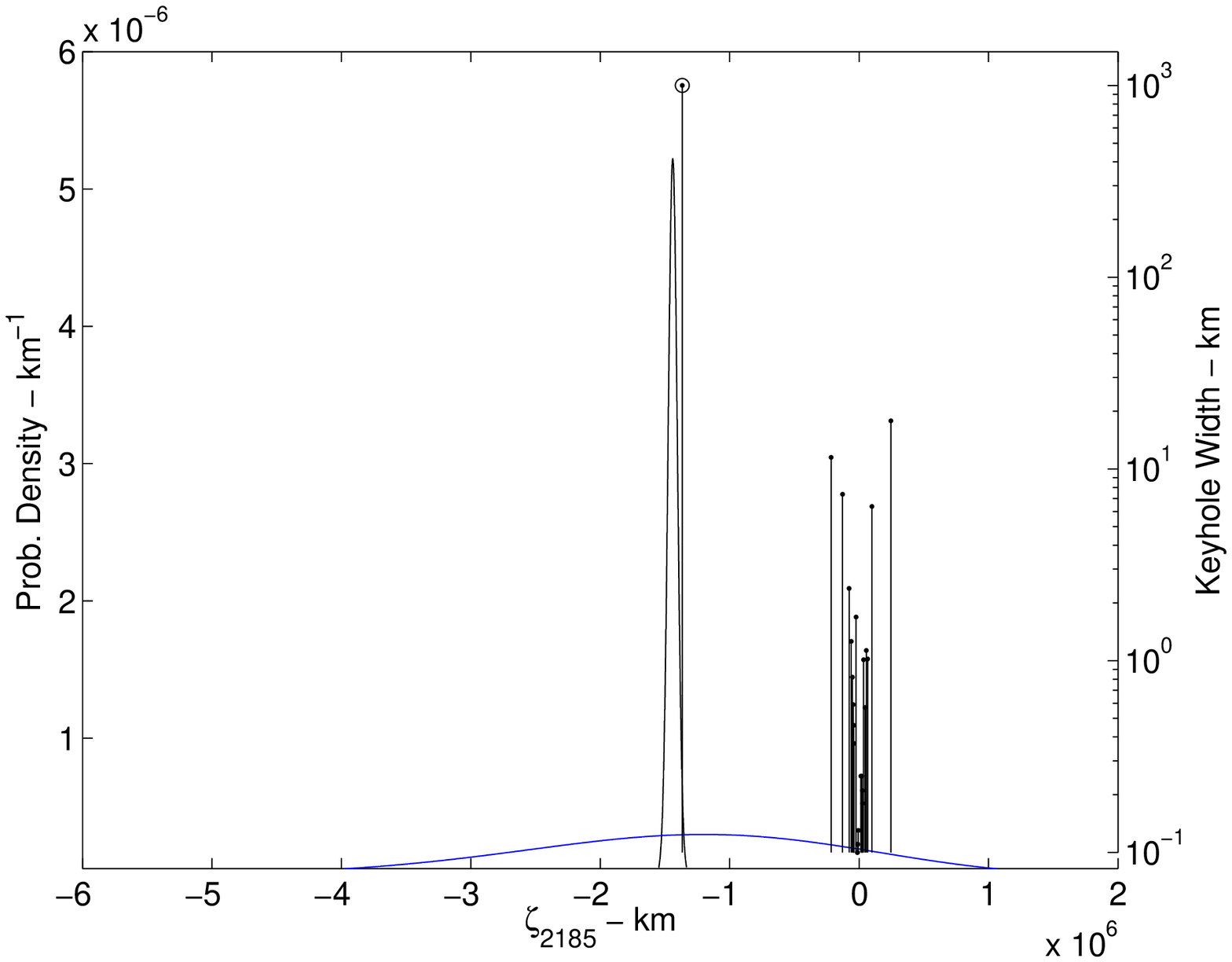}{Two different Probability Density
    Functions (PDF) for the trace of possible solutions on the Target
    Plane of the close approach of asteroid (410777) 2009 FD to the
    Earth in the year 2185. Superimposed and on a different
    vertical scale are the keyholes relative to impacts in
    different years between 2185 and 2196; the height of the bar is
    proportional to the width of the keyhole, thus the Impact
    Probability can be computed as product of the width and the PDF.}
\end{figure}

In Figure~\ref{fig:pdf_yarko_noyarko} we show two probability
distributions, as derived from the orbit determination of the
asteroid (410777) 2009 FD. The narrow peaked distribution corresponds
to an orbit determination with 6 parameters, the initial conditions
only: the standard deviation is $6\times 10^4$ km. The much wider
distribution corresponds to a fit with 7 parameters, including the
constant $A_2$ appearing in the transverse acceleration due to the
Yarkovsky effect: the STD is $\simeq 2.3\times 10^6$ km. The Yarkovsky
effect is a form of non-gravitational perturbation due to thermal
radiation emitted anisotropically by the asteroid, and is indeed very
small. However, when the uncertainty resulting from the covariance
matrix of the orbit determination is propagated for $\sim 170$ years
after the last observation available, not only the Yarkovsky effect
has a long enough time to accumulate but it is also enhanced
by the exponential divergence of nearby orbits, the Lyapounov time
being about $15.3$ years \citep{2009FD}[Figure 5].

The practical consequence of this increase of the uncertainty
arises from the fact that the Target Plane of 2009 FD for 2815
includes some \textit{keyholes}, small portions corresponding to
impacts with the Earth (either at that time or a few years later,
until 2196).  With the 7 parameters solutions these keyholes are
within the range of outcomes with a significant value of the
Probability Density Function, thus the impacts have a non-negligible
probability, the largest being an \textit{Impact Probability} of
$\simeq 1/370$ for 2185.  If on the contrary the orbit was estimated
with 6 parameters only, then the probability would appear to be even
larger for an impact in 2190, and all the other keyholes
(including the one for 2185) would correspond to negligible
impact probabilities. Given that the impact, if it was to occur, would
release an energy equivalent to $3,700$ MegaTons of TNT, this
difference is practically relevant. In fact, the solution including
the Yarkovsky effect leads to a more reliable estimate of the Impact
Probabilities, because the Yarkovsky effect exists and needs to be
taken into account.

Is the discrepancy in the uncertainties with and without the dynamical
parameter in the fit essentially the same phenomenon we have found in
our simple model? We do not know the answer to this question, but we
shall investigate this issue in the future.

\begin{acknowledgements}
This work has been partially supported by the Marie Curie Initial Training
Network Stardust, FP7-PEOPLE-2012-ITN, Grant Agreement 317185.  The
authors were also sponsored by an internal research fund of the
Department of Mathematics of the University of Pisa.
\end{acknowledgements}

\bibliography{family_ages_biblio}

\end{document}